\documentclass[twocol]{ametsocV6.1}

\def\wxch{\textit{WxChallenge}}
\def\UNC{$\textrm{UNC}$}
\def\REL{$\textrm{REL}$}
\def\DSC{$\textrm{DSC}$}

\def\degC{{^{\circ}C}} %For Celsius

\title{A Probabilistic WxChallenge Proposal}

\authors{John R.~Lawson\aff{a,b,*}\correspondingauthor{John Lawson, john.lawson@usu.edu}}

\affiliation{\aff{a}{Department of Mathematics and Statistics, Utah State University, Logan, UT 84322, U.S.A.}\\
\aff{b}{Bingham Research Center, Utah State University, Vernal, UT 84078, U.S.A.}\thanks{* Author's previous affiliation and research location: Department of Geography and Meteorology, Valparaiso University, Valparaiso, IN 46383, U.S.A}}

\abstract{The national forecasting competition \wxch, brainchild of Brad Illston at the University of Oklahoma in 2005, has become a cherished institution played across the United States each year. Participants include students, faculty, alumni, and industry professionals. However, forecasts are given as scalar values without expression of uncertainty, probabilities being a keystone of meteorological forecasting today, and previous attempts to add probabilistic elements to \wxch{} have failed partly due to challenges in making probability forecasting accessible to all, and inability to combine scores with different units while also appropriately rewarding forecasts using proper scoring rules. Much of the competition’s maintenance relies on dedicated volunteers, highlighting need for more automation Hence I propose three new features: (1) automated forecast problems based on morning ensemble guidance, forming prediction baselines, thresholds over which the players demonstrate skill in their later forecast; (2) a spread betting game, where the players allocate 100 confidence credits to the over–under for exceeding a percentile (e.g., 50\%) threshold of a variable (e.g., maximum temperature) derived from the ensemble baseline; and (3) a game where players distribute 100 confidence credits across bins of a continuous variable (e.g., accumulated precipitation) approximating a probability mass function. Forecasts are evaluated using \emph{information} gained over the baseline forecast, yielding additive units of \emph{bits} that allow score combinations of different variables and units. Information gain parallels the Brier Score and is likewise a sound measure of skill due its punishment of hedging. This proposal’s objective is to augment \wxch{} with two new probabilistic games that are accessible, scientifically sound, enjoyable, and optional.}

\begin{document}

%% Necessary!
\maketitle

%%%%%%%%%%%%%%%%%%%%%%%%%%%%%%%%%%%%%%%%%%%%%%%%%%%%%%%%%%%%%%%%%%%%%
% SIGNIFICANCE STATEMENT/CAPSULE SUMMARY
%%%%%%%%%%%%%%%%%%%%%%%%%%%%%%%%%%%%%%%%%%%%%%%%%%%%%%%%%%%%%%%%%%%%%
%
% If you are including an optional significance statement for a journal article or a required capsule summary for BAMS 
% (see www.ametsoc.org/ams/index.cfm/publications/authors/journal-and-bams-authors/formatting-and-manuscript-components for details), 
% please apply the necessary command as shown below:
%
% Significance Statement (all journals except BAMS)
%
%\statement
%	 Enter significance statement here, no more than 120 words. See \url{www.ametsoc.org/index.cfm/ams/publications/author-information/significance-statements/} for details.
%
%% Capsule (BAMS only)
%%
%\capsule
\statement
This study proposes two optional probabilistic games for the \wxch, a national weather-forecasting competition, to improve player familiarity expressing uncertainty in predictions. 

\section*{Notes} This paper was rejected with mixed reviews at the review stage after passing the editorial proposal check at the \textit{Bulletin of the American Meteorological Society.} The author is re-tooling the methodology to address author concerns, but currently, this paper has been withdrawn and is not already available on a preprint server or public repository. No funding was used in the preparation of this manuscript.

\section*{Statement on Generative AI usage} The author used OpenAI's GPT-4 AI service when brainstorming ideas, but did not use generative artificial intelligence for any stage of the manuscript-writing process. Future versions of this statement are subject to change with manuscript-version updates.

%       Enter BAMS capsule here, no more than 30 words. See \url{www.ametsoc.org/index.cfm/ams/publications/author-information/formatting-and-manuscript-components/#capsule} for details.
%
%% * * If using twocol mode, you will need to use the commands "twocolsig" and "twocolcapsule" in place of "sig" and "capsule"
%%      to ensure that the text box correctly spans across both columns.
%

%%%%%%%%%%%%%%%%%%%%%%%%%%%%%%%%%%%%%%%%%%%%%%%%%%%%%%%%%%%%%%%%%%%%%
% MAIN BODY OF PAPER
%%%%%%%%%%%%%%%%%%%%%%%%%%%%%%%%%%%%%%%%%%%%%%%%%%%%%%%%%%%%%%%%%%%%%
%

%% In all cases, if there is only one entry of this type within
%% the higher level heading, use the star form: 
%%
% \section{Section title}
% \subsection*{subsection}
% text...
% \section{Section title}

%vs

% \section{Section title}
% \subsection{subsection one}
% text...
% \subsection{subsection two}
% \section{Section title}

%%%
\section{Introduction}
The national weather-forecasting competition \wxch, having its roots at the University of Oklahoma planted Brad Illston in 2005 and tended by many volunteers since, is a much loved institution played across the United States each year. From the synopsis (\url{http://www.wxchallenge.com/about.php}; accessed 1 January 2023):

\begin{quote}
``The \wxch{} started [...] to create a more dynamic and engaging forecast contest [...] built upon simple forecast submission and rapid forecast verification. [The \wxch] began official operations in the fall of 2006 with 55 participating universities, 41 classrooms, and nearly 1600 participants. Currently, the \wxch{} has been used at over 150 universities, 175 classrooms, and by over 14,000 participants in its short span."
\end{quote}

Participants comprise industry professionals, enthusiasts, and university staff, students, and alumni. Players predict four values for a site in the US over eight 24-h periods, and this continues for Fall and Spring semesters as the forecasting city changes every two weeks of play; the site is selected by the \wxch{} board from a US region determined by player consensus. Players forecast high and low dry-bulb 2-m temperatures to the nearest degree Fahrenheit; maximum 1-min sustained 10-m wind speed to the nearest knot; and rainfall (or snow-water equivalent) accumulation to the nearest hundredth of an inch. For the benefit of readability, I will use the imperial system of units employed by \wxch{} instead of frequent conversions to S.~I.\ units. The familiarity of these units to Americans, and the simplicity of so-called \emph{deterministic} (i.e., certain) forecasts, contribute to the game's accessibility and likely enhance \wxch's popularity and longevity. Nonetheless, best practices at operational weather-forecast offices worldwide often involve expressing uncertainty mathematically (e.g., percentage risk) or linguistically (scenarios; categories) \citep[amongst others]{Mittermaier2013-cz,Hagelin2017-ea,Rothfusz2018-yk,Gascon2019-iv,Roberts2019-ox,Porson2019-bq}. The rapid advent of artificial intelligence (AI), especially large language models such as GPT-4 \citep{Bubeck2023-ls,OpenAI2023-lx} and the open-source Cerebras-GPT \citep{Dey2023-nq}, is likely to yield improved conversion of raw model output to public communication of uncertainty \citep{Solaiman2019-sh,Hagendorff2023-rz}. These models, like many in data analysis, are driven fundamentally by probability theory \citep{Han2022-fc}: it is paramount undergraduate meteorology students have a grasp of risk and uncertainty upon graduation.

Fundamentally, the climate system is a chaotic, complex system \citep{Lorenz1963-zy,Gell-Mann1994-ue} with inherent uncertainty, rendering perfect deterministic forecasts impossible. Further, rapid growth of uncertainties, known as the Butterfly Effect \citep{Palmer2014-cv}, restricts our ability to predict future events with smaller phenomena having a closer predictability \emph{time horizon}. Consider forecasting maximum wind speed on a convective summer's day: the variable distribution of thermals and thunderstorms associated with gusts has a large element of chaotic unpredictability, which necessitates probabilistic forecasting. Predicting a scalar value constitutes a \emph{hedge} (minimizing error through choice of a value close to a perceived mean, instead of issuing one's true belief) to predict maximum wind speed on, e.g., convective days during \wxch. Precise prediction of thunderstorm location and intensity for any given location remains elusive even under the most optimistic of circumstances \citep{Lorenz1963-zy,Durran2016-iv}. This can also be viewed in terms of \emph{information flow}: over time, the usefulness of information about future events diminishes, plateauing to a limit of information gain no greater than random guidance for a similar climate state. This concept of error saturation is akin to an \textbf{information-gain horizon}. A deeper discussion of information theory is presented by \citet{Pierce1980-kt}, \citet{Cover2012-di}, amongst others, while \citet{DelSole2004-lk} and \citet{Weijs2011-cf} summarize applications to meteorology.

This critique of \wxch{} games is not to undermine the educational power vested in many years of player participation in the \wxch, which has garnered much long-lasting appreciation across the U.~S~. Moreover, if the game's annual recurrence aims to encourage students to familiarize themselves with weather phenomena, this goal is met regardless of \wxch's deterministic nature. In 2023, deterministic forecasts persist across mobile-device applications and TV broadcasts, demonstrating a continued demand for binary thinking---for better or worse. However, if a primary educational goal is to prepare students, professionals, and hobbyists for predictions optimized for decision-making in 2023, then incorporating probabilistic elements or expressions of uncertainty is essential. Many researchers and educators have proposed additions (summarized in \citealt{Decker2012-bh} with that author's own probabilistic proposal). Indeed, the \wxch{} administration board has long considered inclusion of probabilistic elements in the competition (Teresa Bals-Elsholz, pers.~comm.), balanced with skepticism that probabilism would be appropriate in \wxch{} (Brad Illston, pers.~comm.), along with interest in reducing the human workload in running the contest. The present proposal acknowledges that \wxch{} serves different roles for various players. Any uncertainty-based extensions must be accessible, considering the diverse educational backgrounds and experiences, while not detracting from the game's enjoyment. Thus I pose: \textbf{can \wxch{} incorporate probabilistic elements without diminishing the enjoyment of participation}?

\section{Background}
Herein, $f$ and $o$ denote forecast and observed probabilities, respectively; e.g., $f = p(o > 21\,\degC) = 0.7$ for a 70\% probability $p$ of observed temperature $o$ to exceed $21\,\degC$. We assume observations are binary (certain).

% We represent the forecast with a vector indicating probabilities of exceedence in the order $\left( \textrm{NO}, \textrm{YES} \right)$. For instance, a probabilistic prediction $\textbf{f} = (1-f, f)$, with boldface denoting a vector, could represent the above probability of exceeding 21\,$\degC$ written as $\textbf{f} = (0.3, 0.7)$. Although we prescribe binary (certain) observations in the current paper, the concept of verification as a vector (i.e., uncertain observations) within the context of information gain is discussed in (Weijs ref).

\subsection{Current \wxch{} Rules}
Full rules are found at \url{www.wxchallenge.org/rules} (accessed 1 January 2023).

\begin{itemize}
    \item One error point is allocated for each \textbf{degree Fahrenheit of dry-bulb temperature} between forecast and observation values; 
    \item Half an error point is allocated for each \textbf{knot of wind magnitude} between forecast and observation;
    \item Precipitation follows an algorithmic rule, with error points allocated thus:
    \begin{itemize}
        \item Trace precipitation is scored as zero;
        \item 0.4 points per 0.01\,in error for the class trace to 0.1 inclusive;
        \item 0.3 points per 0.01\,in error for the class 0.11 to 0.25 inclusive;
        \item 0.2 points per 0.01\,in error for the class 0.26 to 0.5 inclusive;
        \item 0.1 points per 0.01\,in error for the class over 0.5.
    \end{itemize}
\end{itemize}

In this context, observations from the verification METAR sites are assumed to be error-free. The primary scientific issue with these scoring rules is that they are neither standard \citep{Mason2008-vr} nor truly fair \citep{Decker2012-bh}. Further, as these rules are not \emph{proper} (see \emph{desiderata} in \citealt{Benedetti2010-sa}), the game can be hedged, which is an undesired property. For instance, if a player is near the top of a city's rankings on the final day, and their goal is to win that city or bust (``all-in"), they may opt for a bold forecast towards the tails of the notional distribution if they believe other players will play it safe (i.e. hedge, or minimize losses using mean-square error). However, this approach is \emph{reductio ad absurdum}: a player's forecast should not depend on perceived alternative forecasts, lest we reach a scenario where all players assume others' level of caution and no player forecasts their true beliefs. A proper score will optimally reward a forecast that matches the forecaster's confidence in the event happening, not in minimizing error, as analyzed as a time-averaged mean. Some events occur at the tail of the frequency distribution, and an ensemble mean may smudge the ``true" probabilities that may cluster bimodally. Hedging by issuing an ensemble mean, anecdotally a common strategy in \wxch, undermines a forecasting system if it fails to reflect a forecaster's true estimate of the atmospheric uncertainty. This motivates the provision of games for players to express their confidence of a threshold exceedence, and not solely a best guess at a measurement.

% COnsider a table of the three desiderata from Peirolo/Weijs stuff 

\subsection{Practical motivation for change}
Uncertainty is instinctively understood by the layperson when presented in an accessible way \citep{Gigerenzer2005-vx}, which can be challenging (refs.\ therein). Indeed, the recipient of a deterministic weather forecast (e.g., maximum temperature for the coming solar day) typically adds tolerance, even subconsciously. Linguistically, a term such as ``bust" represents failure to pass a threshold in a probabilistic paradigm. A 6-h simulated-reflectivity map from a convection-permitting numerical model showing predicted thunderstorms will likely provide misleading information if the viewer accepts the data with certainty in time and space. However, a layperson's ``mental smoothing" \footnote{This smoothing is likely by default a Gaussian curve in three dimensions, similar to surrogate severe smoothing kernels in \citet{Sobash2016-wo}} is unlikely to reflect the range of potential future states. For example, when one cluster of forecasts call for a thunderstorm outbreak, while another does not, a scenario where guidance clusters into two groups emerges. It is unlikely a layperson would apply appropriate tolerance in such a situation; varying interpretation of risk by the general public is discussed in \citet{Blastland2014-wn}. As humans instinctively default to fast, linear thinking \citep{Kahneman2011-dn,De_Langhe2017-rw,Yama2019-rj}, it is understandable that probabilistic forecasting is challenging to communicate. Probabilistic and \emph{proper} forecasts are crucial for public communication and education \citep[for example]{Krzysztofowicz2001-us} to mitigate extreme weather events \citep{Verbunt2007-op,Roebber2013-ej,Williams2014-zj,Berner2017-ag}. Beyond the scope of this manuscript is the relevance of proper probabilistic forecasts to decision-making regarding \emph{forecast value} and \emph{cost--loss ratio} \citep{Buizza2001-bl}. 

\section{Scoring}
\subsection{Mathematical motivation for change}
The human experience of weather is embedded within the complex, adaptive, chaotic global climate system \citep{Gell-Mann1994-ue}. Before an event occurs, error and uncertainty (e.g., ensemble spread) are indistinguishable: they are discriminated after the fact. There is a limit to prediction of a complex system displaying chaos, and this reduces temporally with the spatial scale. This is intuitive: predicting wind speed at the end of one's garden within seconds is impossible without substantial error. It follows that a 24-h maximum-wind forecast similar to that in the \wxch{} must pair with an expression of uncertainty and/or bounds within which that variable may range.

% We assume observations are perfect ($o \in \{0,1\}$); however, we prohibit forecasts of certainty (0, 1), first due to catastrophic implications of assuming impossiblity/inevitability for an event, and second from the divergence to infinity yielded in Eqn.~\ref{eq:1}.

\subsection{Limitations of the Brier Score}
Mathematically, there are two probabilistic-score families that meet fundamental criteria \citep{Benedetti2010-sa} for scoring rules: Brier- and Information-type scores. The former is based on mean-squared error (MSE), found in the Brier Score (BS; \citealt{Brier1950-lc}), the Ranked Probability Score (RPS; \citealt{Hersbach2000-yb}), and the ensemble Fractions Skill Score (FSS or eFSS; \citealt{Roberts2008-xv,Duc2013-jb}). Information-theoretical concepts are found in ignorance (IGN, \citealt{Roulston2002-eq}), Good's logarithm rule \citep{McCutcheon2019-jq}, information gain \citep{Peirolo2011-sl,Lawson2021-hq,Lawson2024-bu}, and Ranked Ignorance \citep{Todter2012-ou}. In contrast to Brier-type scores, Information-type scores use Kullback-Liebler Divergence ($D_{\textrm{KL}}$), often deployed in artificial-intelligence development as cross-entropy loss \citep{Ramos2018-ze,Bubeck2023-ls}, to optimize prediction that is similar to that herein. This is further compared in \citet{Weijs2010-hg}. Many readers may be unfamiliar with the jargon surrounding information theory; this is an obstacle to the scores gaining traction, along with the increased computational load required by logarithm-based scores (pers.~comm., Harold Brooks and Leonard Smith). The scores are nonetheless similar, as the BS 

\begin{equation}
    \textrm{BS}_t = \sum^{T}_{t=1} (f_t - o_t)^{2}
\end{equation}

is a scaled, second-order Taylor approximation of IGN where the event occurred (i.e., $o = 1$; \citealt{Todter2012-ou}),

\begin{equation}
    \textrm{IGN}_t = \log_{2} f_t
\end{equation}

where the subscript $t$ is for each forecast issued for all samples $T$; and likewise for extensions of the scoring rule. How does the evaluating scientist choose between similar scores? Twofold: first, Brier-type scores have units that match that of the evaluated variable, whereas Information-type scores yield additive \emph{bits}. The advantage is that forecasts of different variables can be summed equally. Second, Brier-type scores diverge from Information-type scores at extreme (\textless1\%) probabilities \citep{Benedetti2010-sa}, and mathematically this divergence to infinity is intractable when calculating and aggregating scores. 

Hence, information gained by a forecast percentage $f$ over a baseline percentage $b$ is measured by information gain (IG):

\begin{equation}\label{eq:ig}
    \textrm{IG}_t = \log_2 \frac{f_t}{b_t}
\end{equation}

This can also be reached by computing IGN for the baseline and player forecasts via logarithmic identities as

\begin{equation}\label{eq:ig_ign}
    \textrm{IG} = \textrm{IGN}_b - \textrm{IGN}_f
\end{equation}

where subscripts $b$ and $f$ denote ignorance in the baseline and forecast probabilities, respectively. (The subscript $t$ is dropped for clarity here.)

\subsection{Advantages of Information Gain and bits}
Outlined above are mathematical reasons to prefer information gain over the Brier Score. There is a philosophical reason for using Information-based scores, also. Information Theory has been described as the biggest innovation of the 20th Century (e.g., \url{https://www.scientificamerican.com/article/claude-e-shannon-founder/}, accessed 1 January 2023), without which we may not have entered the Information Age. Hence, there is an attraction to embrace a framework that unites multiple fields across sciences \citep{DelSole2004-lk,Cover2012-di}. Ultimately, I submit that information theory is the optimal measure of forecast skill. But what do the \emph{bits} represent? If the BS is hamstrung by its dependence on units, IG is initially nebulous in its interpretation without them. This is because Information Theory cares solely about symbols. It is reductionist. The bit is as intrinsic a measurement of information as a gram measures mass.

The Brier-type scores yield scores in units of the variable evaluated (e.g., inches for rainfall); Information-type scores are in \emph{bits}. One bit represents a question; a choice; the uncertainty of a fair coin flip; a level of a decision tree \citep{Williams1997-us}. Entropy represents surprise, where one coin flip contained one bit of uncertainty (entropy) before the event. Hence 2\,bits represent twice the amount of surprise as 1\,bit upon receiving the forecast. The bits yielded by Information-related scores are \emph{additive}, hence we simply add the scores of each prediction no matter the variable. Use of the scores above gives the \wxch{} an opportunity to implement new forecast questions, such as ice accumulation or minimum dew-point temperature, without worry of how to combine scores fairly.

% The receiver would have to be surprised X more times by a non-occurrence of a 1\% event than a non-occurrence for the entropy to be equal 

\subsection{Discrimination and reliability}
The Information-based scores can be decomposed, as with Brier-based scores, into discrimination \DSC, reliability \REL, and uncertainty \UNC. Discrimination resembles inherent goodness, or the ability of the forecast catalog to correctly predict the outcome class (over/under, for instance). Reliability is how well the forecasts are calibrated: for instance, a 20\% forecast of snow should verify for one-in-five of the occasions this probability is issued. Let us neglect \UNC, the inherent uncertainty in the system, as all players are forecasting the same event catalog (i.e., there is no need to normalize a forecast by its base rate). We also have an uncertainty value set by the 1200\,UTC baseline. In the Information-based framework, \DSC{} measures useful information gained by prior detection of the observed event; conversely, \REL{} is information lost in a degree of  ``probabilistic false alarm".

Probabilistic statistics from \wxch{} could be accessed at varying degrees of complexity should players so desire. With sufficient sample size, forecasts evaluated with Information-type scores can be decomposed Brier-style into \DSC{} and \REL. The ability to decompose skill gives the player advice on how to tune the sharpness of their predictions. For instance, consider an average-performing player that has amassed considerable error from spreading points out over many bins: issuance a low confidence in any outcome. Inspecting the \REL--\DSC{} trade-off would inform the player they are being under-confident, on average: a new facet of \wxch{} statistics and learning opportunity. Given the acceleration of AI research, this also familiarizes the player with an analogous issue of the bias--variance tradeoff in AI and its manifestation as over- and under-fitting. High reliability corresponds to low bias, meaning forecast probabilities are near those observed (or binary observations in our case), while high discrimination corresponds to low variance implying a model that gives more concentrated (sharp) predictions.

\section{Creating a baseline to beat}\label{sec:superensemble}
% (cite Ben-Haim book about renaming entropy?
A baseline, known as the \emph{prior} in Bayesian statistics, and \emph{ignorance}, \emph{entropy}, or \emph{uncertainty} in information theory, is established at 1200\,UTC from a superensemble of 36-hour forecasts before the 0000\,UTC deadline for players to submit their predictions. The 12-hour gap between setting the baseline and the submission deadline provides players ample time to submit their forecasts. Using the 1200\,UTC runs of models -- such as the Global Ensemble Forecast System (GEFS, GFS), the North American Model (NAM), and the High-Resolution Rapid Refresh model (HRRR) -- we compute ensemble percentiles (e.g., 50\% for a most-likely case and 90\% for an extreme scenario). Each percentile's corresponding scalar value offers an exceedence threshold for players to make probabilistic prediction on whether that variable (e.g., temperature) will exceed the value corresponding to the percentile, thus categorizing the prediction as ``over" or ``under".

Implementation of automation and verification would be done in Python, while minimal changes are expected for the current site organisation. The two games proposed herein can be considered as beta software, serving as optional side games in which players can choose to participate. Players are free to revert to deterministic forecasting at any point -- for instance, if they opt into the probabilistic games but find them less enjoyable. In such cases, they can predict 100\% to one category for both games discussed below, with the caveat that automation must also bound probabilities to prevent the verification score from diverging to infinity (this is discussed below). 

\subsection{Automation of thresholds}
The game should prioritise excitement for players, which for many weather enthusiasts often includes anomalous events. These players are likely to enjoy considering a high-percentile (e.g., 90\%) threshold exceedence. Additionally, a two-class (over/under) prediction game involving a random\footnote{In a statistical sense.} variable is most challenging when the baseline ratio is 50:50. (In information terms, this is where the event's entropy is maximised.) Therefore, we set a baseline for the player to beat in issuing their 0000\,UTC forecast with the superensemble's 50th and 90th percentiles. For the chosen location, an automated script would perform the following:

\begin{itemize} 
    \item Take the 12--36-h point forecasts from the 1200\,UTC runs of HRRR, NAM, GFS, and GEFS.
    \item Therewith generate a superensemble forecast for the 0000--2400\,UTC \wxch{} period of accumulated precipitation\footnote{or snow-water equivalent.}, maximum and minimum temperature, and maximum sustained wind-speed. 
    \item Pick the 50th and 90th percentile as the "most likely" and "reasonable extreme scenario" representative thresholds for Game 1.
\end{itemize}

Further information on the relationship between probability, forecast verification, and mathematical \emph{surprise} and entropy can be found in summaries of information theory such as \citet{Cover2012-di}.

%\citep{Hagedorn2009-pc}

\section{Game 1: Over/Under}
The first game deploys spread betting (also known as Asian Handicaps due to their popularity in southeast Asia as \emph{hang cheng} bets). Here is an outline of the game:

\begin{itemize}  
    \item Take the 1200\,UTC model baseline for the 50th (ensemble median) and 90th percentiles for the forecast location's variable of interest. These baselines serve as spread-bet over/unders;
    \item After converting each percentile to a value (e.g., 21\,$\degC$), the \wxch{} question becomes: \textit{will the observed variable exceed the thresholds?}, i.e., an over. The player is given 100 confidence credits to assign to either side of the 50th and 90th percentiles' threshold;
    \item We score the prediction $f$ from verification for the event with information gain, using the baseline $b$ as our expected probability (0.5 or 0.1).
\end{itemize}

\begin{figure*}[t]
  \noindent\includegraphics[width=0.89\textwidth,angle=0]{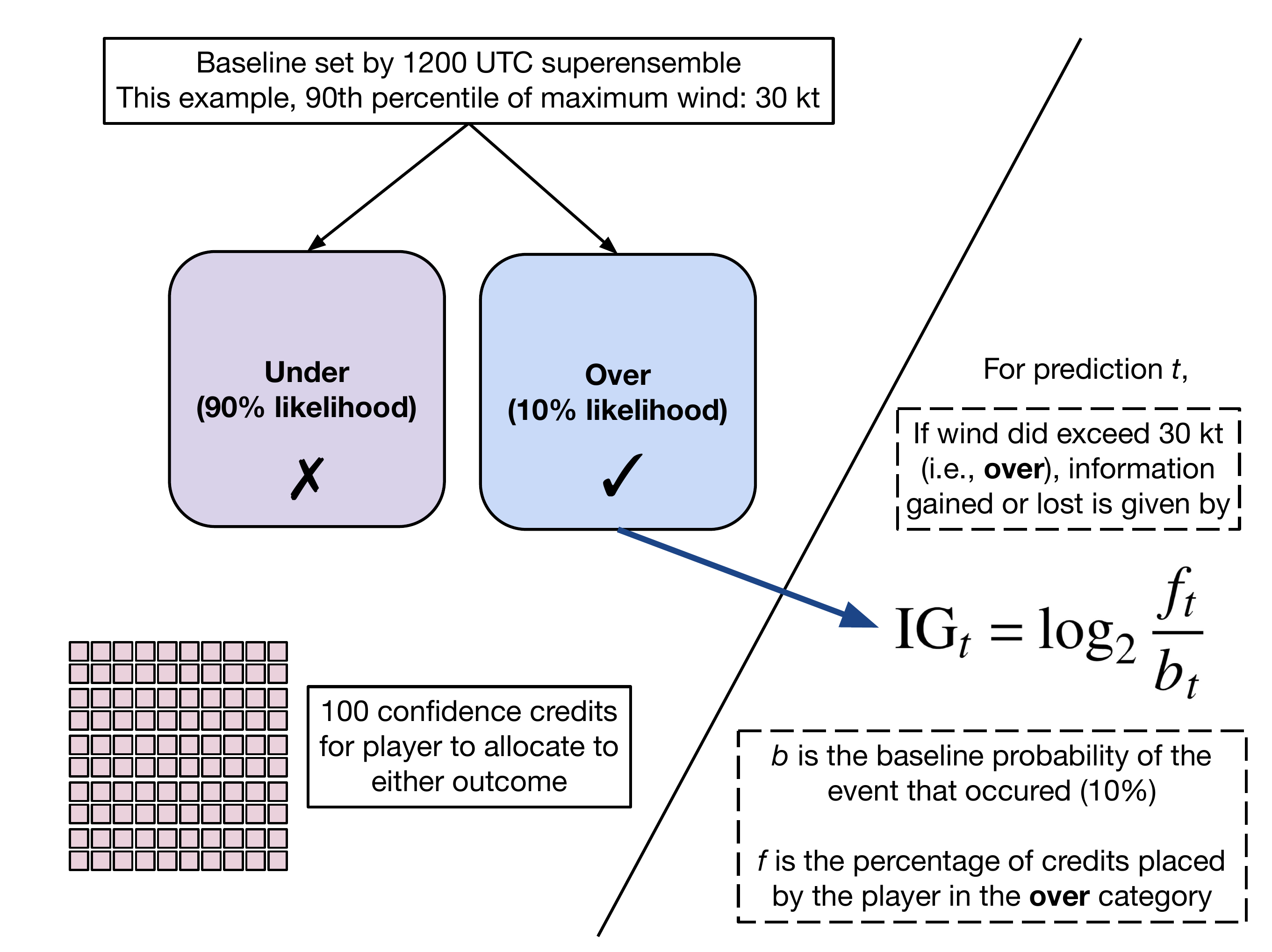}\\
  \caption{Schematic showing the configuration for Game 1 (spread betting). The over/under thresholds are set by the 1200\,UTC superensemble: the 50th and 90th percentiles. The player allocates their 100 confidence credits (bottom left) to the outcome of exceedance or not (central boxes) before 0000\,UTC that evening. With the observation in hand (bottom right), the player is evaluated by measuring the information they gained or lost over the baseline forecast for each percentile. Information gain in \emph{bits} for each event, denoted by subscript $t$, can be mean-averaged over all forecasts and times of all variables' threshold exceedence.}\label{fig:game1}
\end{figure*}

% This is referring to the IG equation - also mention that DKL or XES can
% be simplified when assuming perfect observations.
We evaluate the forecast with information gain (Eqn.~\ref{eq:ig}): predictions are better with increasing information gain for a given forecast. Information is measured in bits, here due to the base-2 logarithm, and their property of additivity means we can mean each information-gain score of each variable. The game schematic is in Fig.~\ref{fig:game1}.

We can interpret one \emph{bit} by further by considering an example of a coin flip. The na\"ive expectation is that of $b = 0.5$; whether it is heads or tails is immaterial. Imagine an coin forecaster who arbitrarily estimates the chance of a heads to be $b = 0.8$. If the heads does appear, the information gained is positive (0.67\,bits, from Eq. 3), as it is surprising and useful to an observer requiring the prediction; otherwise, it is information lost (1.23\,bits), as it has surprised the observer in a negative way. As $b$ moves to either extreme close to zero or unity, the available information to be gained increases, but likewise, the penalty of overconfidence is heightened asymmetrically faster. Hence, given the prior uncertainty is 1\,bit, there is a ($1 - 0.67 = 0.33$) gain or ($1 - 1.23 = 0.23$) loss of information for a good or bad forecast, respectively.

A further forecast challenge may choose a more extreme discriminator, such as 95\%:5\%, which provides a different facet of risk forecasting. Indeed, one may imagine a high-risk, low-frequency endeavor such as in aviation safety where decision-makers must be vigilant against rare, extreme events.

\section{Game 2: Bin Distribution}
\begin{figure*}[t]
  \noindent\includegraphics[width=0.89\textwidth,angle=0]{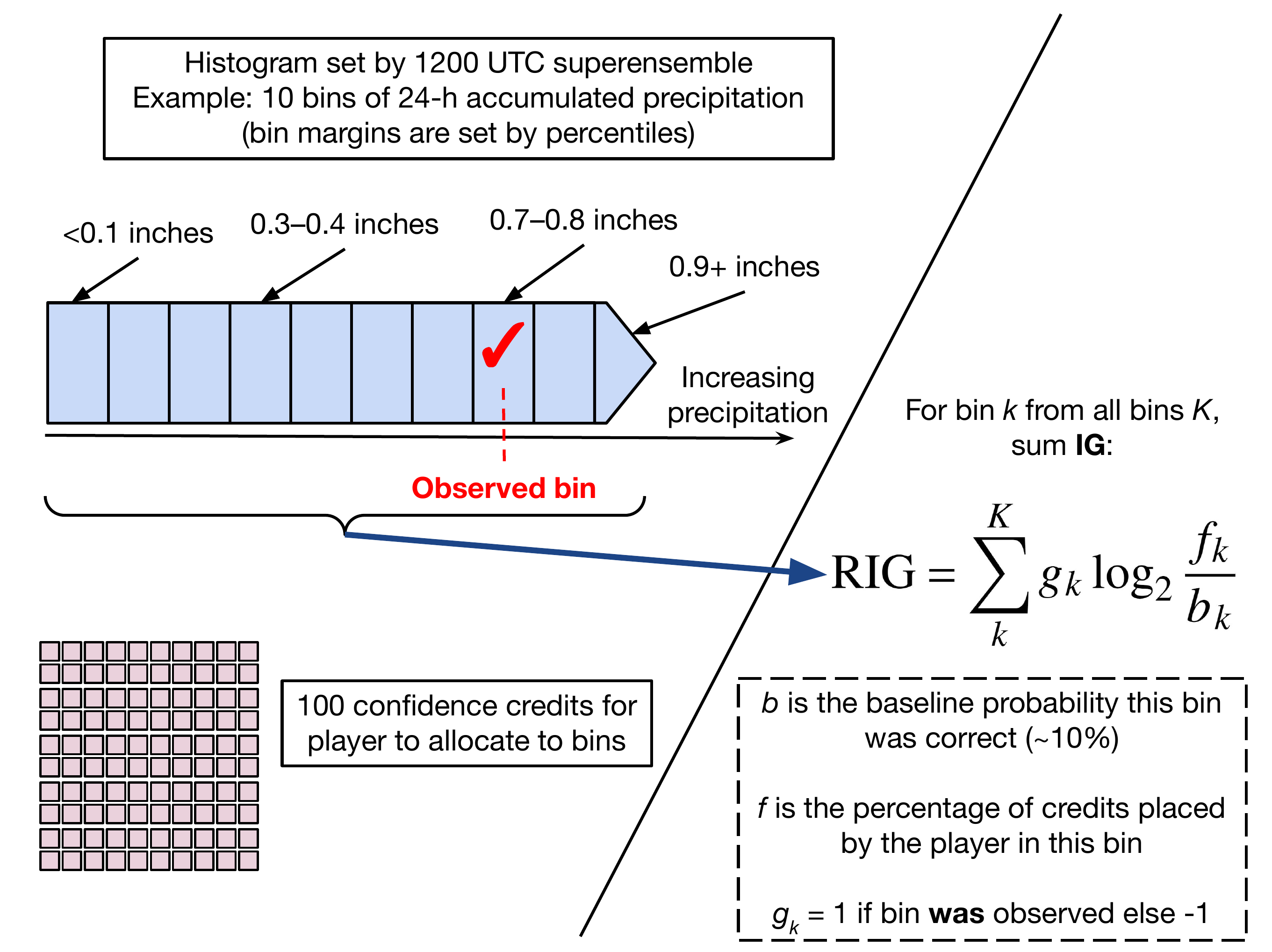}\\
  \caption{Schematic showing the configuration for Game 2 (bin allocation). The bin margins are set by percentile thresholds that relate to sensible weather values (0--0.1\,in, 0.1--0.2\,in, etc). Bins represent equal proportions of the superensemble that predicted each bin's span in sensible-weather units. This span may be non-uniform if the bin-width is modified as the variable $b$ in Eqn.~\ref{eq:rig}. With the observation in hand and considering rounding error, we already know each bin had a $\approx$10\% baseline of occurrence, hence information gained or lost per bin can be summed. With 10 bins, a player putting fewer than 10 credits into a non-observed bin, or \emph{vice versa}, will evaluate as information gained.}\label{fig:game2}
\end{figure*}

We now extend the two-class case of information gain above to a continuous variable such as precipitation accumulation. Naturally, we must choose a modest bin size -- here I choose ten -- to reduce the player burden. From the same superensemble in Section~\ref{sec:superensemble} we can generate a probability mass function (pmf) of exceedence, or percentile values quantised as ten bins/categories (e.g., rainfall as \textless0.2\,in, 0.2--0.4\,in, etc.). This is visualised in Fig.~\ref{fig:game2}. These bins need not be equidistant as they are normalised by their width. Mathematical \emph{surprise} is defined as the difference between a prior and posterior p.m.f. received by a decision-maker: they experience no surprise upon receipt of numerical guidance that does not change their opinion. In Game 2 to simplify this concept, we again give the player 100 confidence credits to assign in any way to the ten bins. Mathematically, this is sampling the pmf distribution with 11 points (this value dictating the so-called \emph{information dimension}), with probabilities allowed from the set $\{ {\underline{p}, 0.01, 0.02... 0.99, \bar{p}} \}$. To avoid divergence to infinity of the logarithm in information scores, a forecasted probability of zero or unity is undesired. We may interpret this as the potentially catastrophic consequence of acting on a certain prediction that did not verify. Hence, we set the maximum and minimum probabilities to $\bar{p}$ and $\underline{p}$, respectively, after which the pmf is rescaled such that all probabilities add to unity (e.g., with a sigmoid or "S-shaped" function). Another choice is using $\underline{p} = \frac{1}{4N}$ as the minimum threshold, where $N$ is the number of superensemble members (creating a quasi-trapezoidal tail to the forecast p.d.f.), with $\bar{p} = 1 - \underline{p}$ as the maximum threshold. The factor of four can be tuned to maximize skill after evaluation.

% Softmax and similar?

Scores are evaluated with an analog of the Ranked Probability Score (RPS; \citealt{Hersbach2000-yb}): its information-theoretical cousin, Ranked Ignorance (RIGN; \citealt{Todter2012-ou}), which is the summation of IGN over all bins, factored by the width of the bin. Herein, we can compute information gained for each class against the baseline using the IG equation, achieved by switching IGN to IG. The bins can be determined by percentiles from previous superensemble runs, converted to absolute values and rounded off for aesthetics and to increase efficiency of calculations. The binned forecasts are then verified using an analogue of RIGN, adapted from Eq. 27 in \citet{Todter2012-ou}: 

\begin{align}\label{eq:rig}
    \textrm{RIG} &= \sum^{K}_{k} g_k \log_2 \frac{f_k}{b_k}
\end{align}

with $f_k$ and $b_k$ denoting the forecast and baseline for bin $k$, respectively, summed over all bins $K$. As each bin is one decile, there is no need to normalize each bin with its width: all bins are equally sized in probability space. Moreover, each bin is weighted by its percentile width via the $b_k$ term, should the bin-width be modified. We determine if information was gained or lost via the factor $g_k \in \{-1,1\}$, where $g_k$ is positive if the bin was observed, else negative. Expanded in full, Eqn.~\ref{eq:rig} shows the maximum score is achieved by placing all 100 credits in the verifying bin. In using ten bins, $b_k \approx 0.1$ for each, but rounding errors are considered implicitly by amending this number. For open-ended variables such as rainfall accumulation, the final bin is  prescriptively weighted the same as the others. The primary drawback of RIG, as with RIGN or RPS, is the sensitivity of the score to the bin size and range. However, I believe ten bins balances how well we approximate the probability distribution versus adding burden on the player.

I stress the above is not a coining of a ``new score" in a field awash with various evaluating scores, but simply a refactoring of RPS with information gain, i.e., deploying IG across a range of bins. Indeed, we can reformulate Eqn.~\ref{eq:rig} as in Eqn.~\ref{eq:ig_ign}, where RIGN from the forecast (ignorance over all bins) is subtracted from that of the baseline.  

In Table~\ref{tab:2x2} we represent each bin's forecast by a 2-by-2 contingency table \citep{Green1966-xl,Jolliffe2003-xs} in a probabilistic framework by using information gained or lost instead of hit/miss to quantify forecast quality. 

\begin{table}[t]\label{tab:2x2}
\caption{Information-gain analogue of the 2-by-2 contingency table. Forecast $f_t$ and baseline $b_t$ probabilities for this sample/time $t$ are verified by binary (certain) observations. When $f_t = b_t$, information transfer is zero.}\label{t1}
\begin{center}
\begin{tabular}{l|cc}
\hline\hline
- & Observed & Not Observed\\
\hline
$f_t < b_t$ & $-\log_2\frac{f_t}{b_t}$ & $\log_2\frac{f_t}{b_t}$ \\
$f_t > b_t$ & $\log_2\frac{f_t}{b_t}$  & $-\log_2\frac{f_t}{b_t}$ \\
\hline
\end{tabular}
\end{center}
\end{table}

\section{Summary}
To encourage introduction of probability forecasts expressing uncertainty into \wxch{} competitions, I have proposed two optional games: one \textbf{assigning confidence credits to the over/under} (spread bet) of exceeding a threshold (such as 24-h maximum temperature); a second \textbf{assigning confidence credits to bins of a continuous variable} (such as accumulated precipitation). Players are scored by providing more useful information than baseline forecasts created using percentiles from a 1200\,UTC superensemble. This is measured with information gain (an analog of the Brier Score) and its ranked form (which considers multiple continuous classes as an analogue of the Ranked Probability Score). Use of information gain yields scores with units of \emph{bits} that are additive, allowing the combination of score from different sensible-weather variables. Information gained from players' predictions can be decomposed into Brier-like components (discrimination and reliability) such that a player can diagnose whether their forecasts are successful more from skillful, sharp confidence (analogous to a calibrated ensemble spread) versus inherent goodness (the ability of a model to correctly forecast a state).

This proposal has a shortcoming that forecast probabilities must not equal to zero or unity to avoid divergence to infinity upon evaluation by Information-based scores. The \wxch{} scripts must bound those values as described above; this process is the prime source of subjectivity by bounding the maximum and minimum of information available to be gained (e.g., \citealt{Lawson2021-hq}). Further, the uncommon deployment of bits as units also requires clarification for many applied meteorologists. 

The larger view begs the questions the author submits may be answered with focus groups, feedback, etc:

\begin{itemize}
    \item What are the goals of the \wxch: for instance, education, entertainment, etc.?
    \item The more taxing the \wxch{} submissions, the more likely a players loses enthusiasm. Where is the balance?
\end{itemize}

With the advent of disruptive \citep{Solaiman2019-sh,Weidinger2021-uf} and powerful artificial intelligence driven by Large Language Models \citep{Bubeck2023-ls,OpenAI2023-lx,Dey2023-nq}, it is more important than ever for meteorologists to understand information and probability theory. Models such as LLMs and neural networks often optimize their model through minimization of cross-entropy loss, which is closely related to information gain. Appreciation of probabilities, and indeed the same information-theoretical roots of verification, prepares the graduate and professional alike for a modern world awash with uncertainty of all genres, and in the presence of quickly encroaching AI breakthroughs.

\subsection{Future work}
This work is a practical implementation from lessons learned from \citet{Lawson2021-hq} and an upcoming manuscript in preparation advocating for information gain when optimizing probabilistic predictions. An anonymized dataset created by \wxch{} can be mined for research; for instance, forecasts for all players could be decomposed into Brier-type reliability (calibration) and discrimination (resolution or sharpness) for insight into systematic player error. 

%%%%%%%%%%%%%%%%%%%%%%%%%%%%%%%%%%%%%%%%%%%%%%%%%%%%%%%%%%%%%%%%%%%%%
% TABLES---INSERT NEAR IN-TEXT DISCUSSION
%%%%%%%%%%%%%%%%%%%%%%%%%%%%%%%%%%%%%%%%%%%%%%%%%%%%%%%%%%%%%%%%%%%%%
%%  Enter tables near where they are discussed within the document. 
%%  Please place tables before/after paragraphs, not within a paragraph.
%%
%
%\begin{table}[t]
%\caption{This is a sample table caption and table layout.  Enter as many tables as
%  necessary at the end of your manuscript. Table from Lorenz (1963).}\label{t1}
%\begin{center}
%\begin{tabular}{ccccrrcrc}
%\hline\hline
%$N$ & $X$ & $Y$ & $Z$\\
%\hline
% 0000 & 0000 & 0010 & 0000 \\
% 0005 & 0004 & 0012 & 0000 \\
% 0010 & 0009 & 0020 & 0000 \\
% 0015 & 0016 & 0036 & 0002 \\
% 0020 & 0030 & 0066 & 0007 \\
% 0025 & 0054 & 0115 & 0024 \\
%\hline
%\end{tabular}
%\end{center}
%\end{table}

%%%%%%%%%%%%%%%%%%%%%%%%%%%%%%%%%%%%%%%%%%%%%%%%%%%%%%%%%%%%%%%%%%%%%
% FIGURES---INSERT NEAR IN-TEXT DISCUSSION
%%%%%%%%%%%%%%%%%%%%%%%%%%%%%%%%%%%%%%%%%%%%%%%%%%%%%%%%%%%%%%%%%%%%%
%%  Enter figures near where they are discussed within the document.
%%  Please place figures before/after paragraphs, not within a paragraph.
% %
%
%\begin{figure}[t]
%  \noindent\includegraphics[width=19pc,angle=0]{figure01.pdf}\\
%  \caption{Enter the caption for your figure here.  Repeat as
%  necessary for each of your figures. Figure from \protect\cite{Knutti2008}.}\label{f1}
%\end{figure}

\clearpage
%%%%%%%%%%%%%%%%%%%%%%%%%%%%%%%%%%%%%%%%%%%%%%%%%%%%%%%%%%%%%%%%%%%%%
% ACKNOWLEDGMENTS
%%%%%%%%%%%%%%%%%%%%%%%%%%%%%%%%%%%%%%%%%%%%%%%%%%%%%%%%%%%%%%%%%%%%%
\acknowledgments
%  Keep acknowledgments (note correct spelling: no ``e'' between the ``g'' and
% ``m'') as brief as possible. In general, acknowledge only direct help in
%  writing or research. Financial support (e.g., grant numbers) for the work done, 
%  for an author, or for the laboratory where the work was performed must be 
%  acknowledged here rather than as footnotes to the title or to an author's name.
%  Contribution numbers (if the work has been published by the author's institution 
%  or organization) should be placed in the acknowledgments rather than as 
%  footnotes to the title or to an author's name.
The author thanks Prof.\ Bart Wolf for his contributions to Game 2 from experience in classes; Prof. Teresa Bals-Elsholz for insight into the WxChallenge board and code configuration; Drs. Kenric Nelson, Dan White, and Corey Potvin for parallel discussion on information theory; and undergraduate meteorology students at Valparaiso University for inspiring conversion. Finally, the author thanks Jimmy Correia for continued discussion after breaking anonymity after the review process.

%%%%%%%%%%%%%%%%%%%%%%%%%%%%%%%%%%%%%%%%%%%%%%%%%%%%%%%%%%%%%%%%%%%%%
% DATA AVAILABILITY STATEMENT
%%%%%%%%%%%%%%%%%%%%%%%%%%%%%%%%%%%%%%%%%%%%%%%%%%%%%%%%%%%%%%%%%%%%%
% 
%
\datastatement The data-availability statement is not applicable in this study.
%  The data availability statement is where authors should describe how the data underlying 
%  the findings within the article can be accessed and reused. Authors should attempt to 
%  provide unrestricted access to all data and materials underlying reported findings. 
%  If data access is restricted, authors must mention this in the statement. See
%  {http://www.ametsoc.org/PubsDataPolicy} for more info.

%%%%%%%%%%%%%%%%%%%%%%%%%%%%%%%%%%%%%%%%%%%%%%%%%%%%%%%%%%%%%%%%%%%%%
% APPENDIXES
%%%%%%%%%%%%%%%%%%%%%%%%%%%%%%%%%%%%%%%%%%%%%%%%%%%%%%%%%%%%%%%%%%%%%
%
%% If only one appendix, use

%\appendix

%% If more than one appendix, use \appendix[<letter>], e.g.,

%\appendix[A] 

%% Appendix title is necessary! For appendix title:

%\appendixtitle{Title of Appendix}

%%% Appendix section numbering (note, skip \section and begin with \subsection)
%
% \subsection{First primary heading}

% \subsubsection{First secondary heading}

% \paragraph{First tertiary heading}

%%%%%%%%%%%%%%%%%%%%%%%%%%%%%%%%%%%%%%%%%%%%%%%%%%%%%%%%%%%%%%%%%%%%%
% REFERENCES
%%%%%%%%%%%%%%%%%%%%%%%%%%%%%%%%%%%%%%%%%%%%%%%%%%%%%%%%%%%%%%%%%%%%%
% Make your BibTeX bibliography by using these commands:
\bibliographystyle{ametsocV6}
\bibliography{paperpile}

\begin{thebibliography}{50}
\providecommand{\natexlab}[1]{#1}
\providecommand{\url}[1]{\texttt{#1}}
\renewcommand{\UrlFont}{\rmfamily}
\providecommand{\urlprefix}{URL }
\expandafter\ifx\csname urlstyle\endcsname\relax
  \providecommand{\doi}[1]{https://doi.org/\discretionary{}{}{}#1}\else
  \providecommand{\doi}{https://doi.org/\discretionary{}{}{}\begingroup
  \urlstyle{rm}\Url}\fi
\providecommand{\eprint}[2][]{\url{#2}}

\bibitem[{Benedetti(2010)}]{Benedetti2010-sa}
Benedetti, R., 2010: Scoring rules for forecast verification. \textit{Mon.
  Weather Rev.}, \textbf{138~(1)}, 203--211.

\bibitem[{Berner et~al.(2017)}]{Berner2017-ag}
Berner, J., and Coauthors, 2017: Stochastic parameterization: Toward a new view
  of weather and climate models. \textit{Bull. Amer. Meteor. Soc.},
  \textbf{98~(3)}, 565--588.

\bibitem[{Blastland and Spiegelhalter(2014)Blastland, and
  Spiegelhalter}]{Blastland2014-wn}
Blastland, M., and D.~Spiegelhalter, 2014: \textit{The Norm Chronicles: Stories
  and Numbers About Danger and Death}. Basic Books.

\bibitem[{Brier(1950)}]{Brier1950-lc}
Brier, G.~W., 1950: Verification of forecasts expressed in terms of
  probability. \textit{Mon. Weather Rev.}, \textbf{78~(1)}, 1--3.

\bibitem[{Bubeck et~al.(2023)}]{Bubeck2023-ls}
Bubeck, S., and Coauthors, 2023: Sparks of artificial general intelligence:
  Early experiments with {GPT-4}. \eprint{2303.12712}.

\bibitem[{Buizza(2001)}]{Buizza2001-bl}
Buizza, R., 2001: Accuracy and potential economic value of categorical and
  probabilistic forecasts of discrete events. \textit{Mon. Weather Rev.},
  \textbf{129~(9)}, 2329--2345.

\bibitem[{Cover and Thomas(2012)Cover, and Thomas}]{Cover2012-di}
Cover, T.~M., and J.~A. Thomas, 2012: \textit{Elements of Information Theory}.
  John Wiley \& Sons.

\bibitem[{de~Langhe et~al.(2017)de~Langhe, Puntoni,, and
  Larrick}]{De_Langhe2017-rw}
de~Langhe, B., S.~Puntoni, and R.~P. Larrick, 2017: Linear thinking in a
  nonlinear world. \textit{Harv. Bus. Rev.}, \textbf{2017~(May-June)}, 11.

\bibitem[{Decker(2012)}]{Decker2012-bh}
Decker, S.~G., 2012: Development and analysis of a probabilistic forecasting
  game for meteorology students. \textit{Bull. Am. Meteorol. Soc.},
  \textbf{93~(12)}, 1833--1843.

\bibitem[{DelSole(2004)}]{DelSole2004-lk}
DelSole, T., 2004: Predictability and information theory. part i: Measures of
  predictability. \textit{J. Atmos. Sci.}, \textbf{61~(20)}, 2425--2440.

\bibitem[{Dey et~al.(2023)}]{Dey2023-nq}
Dey, N., and Coauthors, 2023: {Cerebras-GPT}: Open {Compute-Optimal} language
  models trained on the cerebras {Wafer-Scale} cluster. \eprint{2304.03208}.

\bibitem[{Duc et~al.(2013)Duc, Saito,, and Seko}]{Duc2013-jb}
Duc, L., K.~Saito, and H.~Seko, 2013: Spatial-temporal fractions verification
  for high-resolution ensemble forecasts. \textit{Tellus Ser. A Dyn. Meteorol.
  Oceanogr.}, \textbf{65~(1)}, 18\,171.

\bibitem[{Durran and Weyn(2016)Durran, and Weyn}]{Durran2016-iv}
Durran, D.~R., and J.~A. Weyn, 2016: Thunderstorms do not get butterflies.
  \textit{Bull. Amer. Meteor. Soc.}, \textbf{97~(2)}, 237--243.

\bibitem[{Gasc{\'o}n et~al.(2019)Gasc{\'o}n, Lavers, Hamill, Richardson,
  Ben~Bouall{\`e}gue, Leutbecher,, and Pappenberger}]{Gascon2019-iv}
Gasc{\'o}n, E., D.~Lavers, T.~M. Hamill, D.~S. Richardson,
  Z.~Ben~Bouall{\`e}gue, M.~Leutbecher, and F.~Pappenberger, 2019: Statistical
  post-processing of dual-resolution ensemble precipitation forecasts across
  europe. \textit{Quart. J. Roy. Meteor. Soc.}, \textbf{0~(ja)}.

\bibitem[{Gell-Mann(1994)}]{Gell-Mann1994-ue}
Gell-Mann, M., 1994: Complex adaptive systems. \textit{Complexity: Metaphors,
  Models, and Reality}, Addison-Wesley.

\bibitem[{Gigerenzer et~al.(2005)Gigerenzer, Hertwig, van~den Broek, Fasolo,,
  and Katsikopoulos}]{Gigerenzer2005-vx}
Gigerenzer, G., R.~Hertwig, E.~van~den Broek, B.~Fasolo, and K.~V.
  Katsikopoulos, 2005: ``a 30\% chance of rain tomorrow'': how does the public
  understand probabilistic weather forecasts? \textit{Risk Anal.},
  \textbf{25~(3)}, 623--629.

\bibitem[{Green et~al.(1966)Green, Swets,, and {Others}}]{Green1966-xl}
Green, D.~M., J.~A. Swets, and {Others}, 1966: \textit{Signal detection theory
  and psychophysics}, Vol.~1. Wiley New York.

\bibitem[{Hagelin et~al.(2017)Hagelin, Son, Swinbank, McCabe, Roberts,, and
  Tennant}]{Hagelin2017-ea}
Hagelin, S., J.~Son, R.~Swinbank, A.~McCabe, N.~Roberts, and W.~Tennant, 2017:
  The met office convective-scale ensemble, {MOGREPS-UK}. \textit{Quart. J.
  Roy. Meteor. Soc.}, \textbf{143~(708)}, 2846--2861.

\bibitem[{Hagendorff(2023)}]{Hagendorff2023-rz}
Hagendorff, T., 2023: Machine psychology: Investigating emergent capabilities
  and behavior in large language models using psychological methods.
  \eprint{2303.13988}.

\bibitem[{Han et~al.(2022)Han, Pei,, and Tong}]{Han2022-fc}
Han, J., J.~Pei, and H.~Tong, 2022: \textit{Data Mining: Concepts and
  Techniques}. Morgan Kaufmann.

\bibitem[{Hersbach(2000)}]{Hersbach2000-yb}
Hersbach, H., 2000: Decomposition of the continuous ranked probability score
  for ensemble prediction systems. \textit{Weather Forecast.}, \textbf{15~(5)},
  559--570.

\bibitem[{Jolliffe and Stephenson(2003)Jolliffe, and
  Stephenson}]{Jolliffe2003-xs}
Jolliffe, I.~T., and D.~B. Stephenson, 2003: \textit{Forecast Verification: A
  Practitioner's Guide in Atmospheric Science}. John Wiley \& Sons.

\bibitem[{Kahneman(2011)}]{Kahneman2011-dn}
Kahneman, D., 2011: \textit{Thinking, Fast and Slow}. Macmillan.

\bibitem[{Krzysztofowicz(2001)}]{Krzysztofowicz2001-us}
Krzysztofowicz, R., 2001: The case for probabilistic forecasting in hydrology.
  \textit{J. Hydrol.}, \textbf{249~(1)}, 2--9.

\bibitem[{Lawson et~al.(2021)Lawson, Potvin, Skinner,, and
  Reinhart}]{Lawson2021-hq}
Lawson, J.~R., C.~K. Potvin, P.~S. Skinner, and A.~E. Reinhart, 2021: The vice
  and virtue of increased horizontal resolution in ensemble forecasts of
  tornadic thunderstorms in {low-CAPE}, high-shear environments. \textit{Mon.
  Weather Rev.}, \textbf{149~(4)}, 921--944.

\bibitem[{Lorenz(1963)}]{Lorenz1963-zy}
Lorenz, E.~N., 1963: Deterministic nonperiodic flow. \textit{J. Atmos. Sci.},
  \textbf{20}, 130--141.

\bibitem[{Mason(2008)}]{Mason2008-vr}
Mason, S.~J., 2008: Understanding forecast verification statistics.
  \textit{Met. Apps}, \textbf{15~(1)}, 31--40.

\bibitem[{McCutcheon(2019)}]{McCutcheon2019-jq}
McCutcheon, R.~G., 2019: In favor of logarithmic scoring. \textit{Philos.
  Sci.}, \textbf{86~(2)}, 286--303.

\bibitem[{Mittermaier(2013)}]{Mittermaier2013-cz}
Mittermaier, M.~P., 2013: A strategy for verifying {Near-Convection-Resolving}
  model forecasts at observing sites. \textit{Weather Forecast.},
  \textbf{29~(2)}, 185--204.

\bibitem[{{OpenAI}(2023)}]{OpenAI2023-lx}
{OpenAI}, 2023: {GPT-4}. Tech. rep.

\bibitem[{Palmer et~al.(2014)Palmer, D{\"o}ring,, and Seregin}]{Palmer2014-cv}
Palmer, T.~N., A.~D{\"o}ring, and G.~Seregin, 2014: The real butterfly effect.
  \textit{Nonlinearity}, \textbf{27~(9)}, R123.

\bibitem[{Peirolo(2011)}]{Peirolo2011-sl}
Peirolo, R., 2011: Information gain as a score for probabilistic forecasts.
  \textit{Met. Apps}, \textbf{18~(1)}, 9--17.

\bibitem[{Pierce(1980)}]{Pierce1980-kt}
Pierce, J.~R., 1980: \textit{An Introduction to Information Theory: Symbols,
  Signals and Noise}. Dover Publications.

\bibitem[{Porson et~al.(2019)Porson, Hagelin, Boyd, Roberts, North, Webster,,
  and Jeff~Chun‐Fung}]{Porson2019-bq}
Porson, A.~N., S.~Hagelin, D.~F.~A. Boyd, N.~M. Roberts, R.~North, S.~Webster,
  and L.~O. Jeff~Chun‐Fung, 2019: Extreme rainfall sensitivity in
  convective‐scale ensemble modelling over singapore. \textit{Quart. J. Roy.
  Meteor. Soc.}

\bibitem[{Ramos et~al.(2018)Ramos, Franco-Pedroso, Lozano-Diez,, and
  Gonzalez-Rodriguez}]{Ramos2018-ze}
Ramos, D., J.~Franco-Pedroso, A.~Lozano-Diez, and J.~Gonzalez-Rodriguez, 2018:
  Deconstructing {Cross-Entropy} for probabilistic binary classifiers.
  \textit{Entropy}, \textbf{20~(3)}.

\bibitem[{Roberts et~al.(2019)Roberts, Jirak, Clark, Weiss,, and
  Kain}]{Roberts2019-ox}
Roberts, B., I.~L. Jirak, A.~J. Clark, S.~J. Weiss, and J.~S. Kain, 2019:
  {PostProcessing} and visualization techniques for {Convection-Allowing}
  ensembles. \textit{Bull. Am. Meteorol. Soc.}, \textbf{100~(7)}, 1245--1258.

\bibitem[{Roberts and Lean(2008)Roberts, and Lean}]{Roberts2008-xv}
Roberts, N.~M., and H.~W. Lean, 2008: {Scale-Selective} verification of
  rainfall accumulations from high-resolution forecasts of convective events.
  \textit{Mon. Weather Rev.}, \textbf{136~(1)}, 78--97.

\bibitem[{Roebber(2013)}]{Roebber2013-ej}
Roebber, P.~J., 2013: Using evolutionary programming to generate skillful
  extreme value probabilistic forecasts. \textit{Mon. Weather Rev.},
  \textbf{141~(9)}, 3170--3185.

\bibitem[{Rothfusz et~al.(2018)Rothfusz, Schneider, Novak, Klockow-McClain,
  Gerard, Karstens, Stumpf,, and Smith}]{Rothfusz2018-yk}
Rothfusz, L.~P., R.~Schneider, D.~Novak, K.~Klockow-McClain, A.~E. Gerard,
  C.~Karstens, G.~J. Stumpf, and T.~M. Smith, 2018: {FACETs}: A proposed
  {Next-Generation} paradigm for {High-Impact} weather forecasting.
  \textit{Bull. Am. Meteorol. Soc.}, \textbf{99~(10)}, 2025--2043.

\bibitem[{Roulston and Smith(2002)Roulston, and Smith}]{Roulston2002-eq}
Roulston, M.~S., and L.~A. Smith, 2002: Evaluating probabilistic forecasts
  using information theory. \textit{Mon. Weather Rev.}, \textbf{130~(6)},
  1653--1660.

\bibitem[{Sobash et~al.(2016)Sobash, Schwartz, Romine, Fossell,, and
  Weisman}]{Sobash2016-wo}
Sobash, R.~A., C.~S. Schwartz, G.~S. Romine, K.~R. Fossell, and M.~L. Weisman,
  2016: Severe weather prediction using storm surrogates from an ensemble
  forecasting system. \textit{Weather Forecast.}, \textbf{31~(1)}, 255--271.

\bibitem[{Solaiman et~al.(2019)}]{Solaiman2019-sh}
Solaiman, I., and Coauthors, 2019: Release strategies and the social impacts of
  language models. \eprint{1908.09203}.

\bibitem[{T{\"o}dter and Ahrens(2012)T{\"o}dter, and Ahrens}]{Todter2012-ou}
T{\"o}dter, J., and B.~Ahrens, 2012: Generalization of the ignorance score:
  Continuous ranked version and its decomposition. \textit{Mon. Weather Rev.},
  \textbf{140~(6)}, 2005--2017.

\bibitem[{Verbunt et~al.(2007)Verbunt, Walser, Gurtz, Montani,, and
  Sch{\"a}r}]{Verbunt2007-op}
Verbunt, M., A.~Walser, J.~Gurtz, A.~Montani, and C.~Sch{\"a}r, 2007:
  Probabilistic flood forecasting with a {Limited-Area} ensemble prediction
  system: Selected case studies. \textit{J. Hydrometeorol.}, \textbf{8~(4)},
  897--909.

\bibitem[{Weidinger et~al.(2021)}]{Weidinger2021-uf}
Weidinger, L., and Coauthors, 2021: Ethical and social risks of harm from
  language models. \eprint{2112.04359}.

\bibitem[{Weijs and van~de Giesen(2011)Weijs, and van~de Giesen}]{Weijs2011-cf}
Weijs, S.~V., and N.~van~de Giesen, 2011: Accounting for observational
  uncertainty in forecast verification: An {Information-Theoretical} view on
  forecasts, observations, and truth. \textit{Mon. Weather Rev.},
  \textbf{139~(7)}, 2156--2162.

\bibitem[{Weijs et~al.(2010)Weijs, van Nooijen,, and van~de
  Giesen}]{Weijs2010-hg}
Weijs, S.~V., R.~van Nooijen, and N.~van~de Giesen, 2010: {Kullback--Leibler}
  divergence as a forecast skill score with classic
  {Reliability--Resolution--Uncertainty} decomposition. \textit{Mon. Weather
  Rev.}, \textbf{138~(9)}, 3387--3399.

\bibitem[{Williams(1997)}]{Williams1997-us}
Williams, G.~P., 1997: \textit{Chaos theory tamed}. Joseph Henry Press.

\bibitem[{Williams et~al.(2014)Williams, Ferro,, and
  Kwasniok}]{Williams2014-zj}
Williams, R.~M., C.~A.~T. Ferro, and F.~Kwasniok, 2014: A comparison of
  ensemble post-processing methods for extreme events. \textit{Q.J.R. Meteorol.
  Soc.}, \textbf{140~(680)}, 1112--1120.

\bibitem[{Yama and Zakaria(2019)Yama, and Zakaria}]{Yama2019-rj}
Yama, H., and N.~Zakaria, 2019: Explanations for cultural differences in
  thinking: Easterners' dialectical thinking and westerners' linear thinking.
  \textit{J. Cogn. Psychol.}, \textbf{31~(4)}, 487--506.

\end{thebibliography}

\end{document}